\documentclass[aps,reprint,amsmath,amssymb]{revtex4-1}
\usepackage{hyperref}
\usepackage{graphicx}
\def\slash#1{\setbox0=\hbox{$#1$}#1\hskip-\wd0\hbox to\wd0{\hss\sl/\/\hss}}
\usepackage{amsmath}
\usepackage{hyperref}
\usepackage{slashed}
\usepackage{amsfonts}
\usepackage{graphicx}
\usepackage{ulem}
\usepackage{xcolor}
\newcommand\redout{\bgroup\markoverwith{\textcolor{red}{\rule[0.5ex]{6pt}{1.0pt}}}\ULon}

\begin{document}
\bibliographystyle{plain}
\title{Phenomenological Implications of Very Special Relativity}

\author{Alekha C. Nayak}
\email{acnayak@iitk.ac.in}
\author{Pankaj Jain}
\email{pkjain@iitk.ac.in}

\affiliation{Department of Physics, Indian Institute of Technology, Kanpur\\ Kanpur,
India- 208016}
\begin{abstract}
We discuss several phenomenological implications of  
 Very Special Relativity (VSR). It is
assumed that there is a small violation of Lorentz invariance and the 
true symmetry group of nature is a 
subgroup called SIM(2). 
This symmetry group postulates the existence of a fundamental or preferred direction in space-time. We study its implications by using an effective 
action which violates Lorentz invariance but respects VSR. We find
that the problem of finding the masses of fundamental fermions is in general   
intractable in the presence of VSR term. The problem can be solved
only in special cases which we pursue in this paper. We next determine
the signal of VSR in torsion pendulum experiment as well as clock 
comparison experiment. We find that VSR predicts a signal which is different
from other Lorentz violating theories and hence a dedicated data analysis
is needed in order to impose reliable limits. Assuming that signal is
absent in data we determine the limits that can be imposed on the VSR 
parameters. We also study the implications of VSR  
in particle decay experiments taking the charged pion and kaon 
decay as an example.
The effective interaction between the charged pion and the
 final state leptons is related to the
fundamental VSR mass terms through a loop calculation. 
We also predict a shift in the angular dependence
of the decay products due to VSR. In particular we find that these 
no longer display azimuthal symmetry with respect
to the momentum of the pion. Furthermore the azimuthal and polar angle
distributions show time dependence with a period of a sidereal day.
This time dependence provides us with a novel method to test VSR in future
experiments.

\end{abstract}

\maketitle
\section{Introduction} 
Lorentz invariance is experimentally verified to a very high degree of accuracy. Nevertheless, it is interesting to consider
models which postulate a small violation of this symmetry. In particular, 
many quantum gravity models predict breaking of Lorentz invariance at Planck scale energy 
($M_{Pl} \approx 10^{19}$ GeV) \cite{Colladay:1998fq}. 
It is rather interesting that the observational data already rules out most of these models, except those based on
supersymmetry \citep{Collins:2004bp,GrootNibbelink:2004za,Jain:2005as,Polchinski:2011za}. In such models, violation of Lorentz
invariance is suppressed by the factor  
$\frac{M_{SUSY}^2}{M_{Pl}^2}$ \citep{Jain:2005as}. 
In these models 
the effects of Lorentz violation (LV) grow with energy
and are significant only
at very high energies. 

An alternative framework to implement violation of Lorentz invariance is
provided by Very Special Relativity (VSR) \cite{Cohen:2006ky}. 
In this framework one
postulates that only a subgroup, such as, T(2), E(2), HOM(2) and SIM(2),
of the full Lorentz group remains preserved \cite{Cohen:2006ky}. 
The generators of  HOM(2), for example,  are $T_1=K_x+J_y$,  $T_2
=K_y-J_x$ and $K_z$ where $\bold{J}$ and $\bold{K}$ represent rotation and 
boost respectively, while those of SIM(2) are $T_1$, $T_2$, $J_z$ and $K_z$. 
A theory which is invariant only under one of these subgroups but
not the full Lorentz group, necessarily breaks the discrete symmetries
 P, T, CP (or CT). However 
the dispersion relations of particles remain unchanged. Hence several 
consequence of SR, such as frame invariance of the speed of light,
 time dilation and velocity addition remain preserved \cite{Cohen:2006ky,Das:2009fi}. 
This also implies that some of the standard
high energy tests of LV are not applicable in this case.  

It is useful to define a null vector
\begin{equation}
 n^{\mu}=(1,0,0,1)\,, 
\label{eq:vectorn}
\end{equation}
which is invariant under E(2) and T(2) transformations
but not under HOM(2) and SIM(2). 
In this paper we shall primarily be interested in small violations of 
Lorentz invariance which preserve SIM(2). We shall implement this by
using effective Lagrangian approach and construct interaction terms
in terms of $n^\mu$ which 
respect SIM(2) but violate Lorentz invariance. 
The vector $n^\mu$ is given by Eq. \ref{eq:vectorn} only in a particular
reference frame. 
In general, the form of $n^\mu$ would change
under Lorentz transformations and rotations. 
However it is always possible to make a HOM(2) (and SIM(2)) 
transformation into the
rest frame of a particle \cite{Cohen:2006ky}. 
Under these transformations $n^\mu$ changes at most by an overall factor
which cancels out in the calculation of decay rates.
Hence we can choose
a frame at rest with respect to the particle or to the laboratory in which
$n^\mu$ takes the form given in Eq. \ref{eq:vectorn}. However the
orientation of the particle momentum relative to the z-axis in this
frame has to be taken into account 
while making experimental predictions, as discussed below.  

There has been considerable theoretical effort in order to understand 
the phenomenological implications of VSR 
\cite{Cohen:2006ir,Dunn:2006,Fan:2006nd,Cohen:2006sc,Gibbons:2007iu,Bernardini:2007ex,Lee:2015tcc,SheikhJabbari:2008nc,Ahluwalia:2010zn,2012PhRvD..85j5009V,Alfaro:2013uva,Cheon:2009zx,Alfaro:2015fha}.
In this paper we illustrate some phenomenological implications of VSR 
 \cite{Cohen:2006ky}
using an effective action approach.
We assume that Lorentz violating, VSR effects are small and can be treated 
perturbatively. We add an effective, gauge invariant VSR invariant
mass terms for leptons and quarks to the Standard Model (SM) action.  
Such a mass term is interesting since it can potentially explain the 
neutrino masses and mixings without requiring a right handed neutrino.
However a detailed analysis of the resulting model is so far lacking
in the literature. As we argue in section
\ref{sec:effective L}, the model in the general case
becomes rather intractable and leads to a mathematical structure 
incompatible with quantum mechanics. 
Hence we are unable to make reliable predictions in the general case
and impose some constraints on the parameter space in order make
the problem solvable. We next consider limits that can be imposed 
 on the restricted set of VSR
parameters using torsion pendulum \cite{Heckel:2006} 
and clock comparison experiments \cite{Cane:2004}.
These can impose limits on the VSR contributions to the electron and
nucleon masses respectively. The latter can be used to constrain the VSR up 
and down quark masses. We determine the time dependence of the  
 signal that VSR produces in such experiments due to rotation of the Earth.  
We find that the signal is different from what is expected in a generic 
LV theory and requires a dedicated data analysis in order to impose 
proper limits. We determine the level at which the electron and nucleon
masses can be constrained in such experiments. 
 
We also study the implications of VSR for
 elementary particle decay experiments taking the charged
pion and kaon decays as an example. Using the uncertainty in
the observed decay rates we impose a limit on the VSR contribution
to the up, down and strange quark masses.  
We also show that VSR leads to anisotropic distribution of decay products in
the pion (or kaon) rest frame. 
Furthermore it leads to azimuthal angle dependence
in the laboratory frame. The differential cross section also picks up 
time dependence due to rotation of the Earth.
 Similar effects are likely arise in
a wide range of decay and scattering processes within VSR. 
Such effects have so far not been studied within the framework
of VSR although some studies have been performed in other Lorentz violating theories \cite{Garg:2011aa}.
As we have already mentioned, in the latter case the effect is likely to be seen only
at very high energies whereas the effects associated with VSR may be observable at
low energies.
Charged pion decay has been studied to constrain 
LV in the weak sector  \cite{Altschul:2013yja,Noordmans:2014bua}. 
 However none of these studies have investigated this
decay process within the framework of VSR.

\section{VSR invariant effective Lagrangian}
\label{sec:effective L}

We work within the framework of a generalized 
SM in which the Lorentz violating terms
which respect VSR are introduced 
using the effective action approach. The corresponding Lagrangian
density can be written as
\begin{equation}
{\cal L} = {\cal L}_g + {\cal L}_Y
\label{eq:model}
\end{equation}
where we have split the terms into the gauge and the Yukawa sector. 
The gauge terms for the case of leptons can be written as 
\begin{eqnarray}
{\cal L}_g &=& 
i \left(\begin{array}{cc}
\bar\nu_{i} & \bar e_{i}
\end{array}\right)_L 
\left(\slashed{\partial} + i{g'\over 2}\slashed{A} - i{g\over 2}
\tau\cdot \slashed{W}\right)
\left(\begin{array}{c}
\nu_{i}\\   e_{i}
\end{array}\right)_L \nonumber\\ 
&+& i\bar e_{iR}\left(\slashed{\partial} + ig'\slashed{A}\right)e_{iR}
+{\cal L}_{VSR} 
\label{eq:leptons}
\end{eqnarray}
where $i$ is the family index and $A_\mu$ and $W^a_\mu$ represent the U(1)
and SU(2) gauge fields. The corresponding
 gauge couplings are denoted by $g_1$ and
$g_2$ respectively. The Lorentz violating, VSR invariant term can be 
expressed as
\begin{eqnarray}
{\cal L}_{VSR} &=&  
{i\over 2}
\left(\begin{array}{cc}
\bar\nu_{i} & \bar e_{i}
\end{array}\right)_L 
\left[{\tilde M_L}^2\right]_{ij}
{\slashed{n}\over n\cdot D}
\left(\begin{array}{c}
\nu_{j}\\   e_{j}
\end{array}\right)_L\nonumber\\
&+&  {i\over 2} \bar e_{iR} \left[{\tilde M_R}^2\right]_{ij}
{\slashed{n}\over n\cdot D} e_{jR}
\label{eq:VSRmodel}
\end{eqnarray}
where $n^\mu$ is the null vector defined in
Eq. \ref{eq:vectorn}. 
Here we shall assume that neutrinos
do not acquire any mass terms other than those arising out of VSR.
The mass matrices $\tilde M_L^2$ and $\tilde M_R^2$ need not be diagonal 
but have to be Hermitian. We could diagonalize them by a unitary transformation
but then the standard mass terms for charged leptons generated through the  
Yukawa interactions will necessarily be non-diagonal. 
After expanding the Higgs field ${\cal H}$ around its vacuum expectation value (vev) $v$,
the Yukawa terms yield
\begin{equation}
{\cal L}_Y = 
- \bar e_{iL}
 M_{_{ij}} 
   \, e_{jR}
+ \left(\begin{array}{cc}
\bar\nu_{i} & \bar e_{i}
\end{array}\right)_L g_{_{Yij}}\tilde{\cal H} 
   \, e_{jR}
+h.c.
\label{eq:Yukawa1}
\end{equation}
where the mass matrix $M=-g_Yv/\sqrt{2}$\,, $g_Y$ is the Yukawa coupling matrix
 and $\tilde{\cal H} $ represent the
fluctuations of the Higgs field around its vev. 

We next diagonalize the mass matrix $M$ by the transformation
\begin{equation}
  e_{iL} 
\rightarrow
U_{Lij}  e_{jL}\, \,\,\,,\,\,\, e_{iR}\rightarrow
U_{Rij} e_{jR} 
\label{eq:unitarytrans}
\end{equation}
where $U_L$ and $U_R$ are unitary matrices. 
Furthermore we diagonalize the VSR neutrino $\tilde M_L^2$ by the 
transformation 
\begin{equation}
  \nu_{iL} \rightarrow V_{Lij}  \nu_{jL}
\label{eq:nuunitarytrans}
\end{equation}
The VSR charged lepton mass terms can now be written as  
\cite{Dunn:2006,Alfaro:2015fha}
\begin{eqnarray}
{\cal L}_{VSR} &=&  
{i\over 2}\bar\nu_{iL}M_{\nu ij}^2 {\slashed{n}\over n\cdot D}\nu_{jL}   
+  
{i\over 2}\bar e_{iL} \left[ M_L^2 \right]_{ij} 
{\slashed{n}\over n\cdot D}e_{jL}  \nonumber \\ 
&+&
{i\over 2} \bar e_{iR} \left[M_R^2 \right]_{ij}
{\slashed{n}\over n\cdot D} e_{jR }
\label{eq:VSRmodel1}
\end{eqnarray}
where $M^2_\nu$ is a diagonal matrix and $M_L^2=U_e^\dagger M_\nu^2 U_e$
and $M_R^2= U_R^\dagger\tilde M_R^2 U_R$ are non-diagonal charged lepton
mass matrices. Here $U_e = U^\dagger_LV$ is the neutrino mixing matrix. 
The resulting Lagrangian nicely explains the neutrino masses and mixings
but considerably complicates the propagation of charged leptons. The
charged lepton Dirac equation gets modified to 
\begin{eqnarray}
\left[\slashed p - M_D - {1\over 2} M_+^2 {\slashed{n}\over n\cdot p} 
- {1\over 2} M_-^2 {\slashed{n}\gamma_5\over n\cdot p}\right]\psi
= 0
\label{eq:modifiedDirac}
\end{eqnarray}
where $M_D$ is the diagonal Dirac mass matrix,
$$
\psi = 
\left(\begin{array}{c}
e\\  \mu \\\tau 
\end{array}\right)
$$
is a 12 component lepton multiplet with $e$, $\mu$ and $\tau$ representing
the 4 component Dirac spinors for these leptons and $M_\pm^2 = (M_L^2 \pm M_R^2)/2$. 

The matrix $M_L^2$ is fixed by the neutrino masses and mixings whereas
$M_R^2$ is completely unknown. Hence, excluding some special cancellations,
 we expect that in general both
$M_\pm^2$ would be non-zero and non-diagonal. This makes 
 Eq. \ref{eq:modifiedDirac} rather complicated and untractable since
it leads to mixing both between different spinors as well
as between families. Furthermore it does not even lead to a Hamiltonian
structure. We see this by going to the non-relativistic limit
and setting the three momentum $\vec p=0$.  
In this limit the equation can be written as
\begin{equation}
H\psi = E\psi
\label{eq:Sch}
\end{equation}
where $H$ is the generalized Dirac Hamiltonian
\begin{eqnarray}
H =   \gamma^0M_D + {M_+^2\over 2E} \Gamma 
+ {M_-^2\over 2E}  \Gamma \gamma^5\,,
\label{eq:genDiracH}
\end{eqnarray}
$\Gamma=1-\gamma^0\gamma^3$.
Let us first consider the simpler case in which 
 the matrices $M_+^2$ and $M_-^2$ are diagonal. In this case we 
can treat each generation of fermions independently. Hence we focus
on a single four component spinor and set $M_D$, $M_+^2$ and $M_-^2$ equal 
to their corresponding diagonal entries $m_D$, $m_+^2$ and $m_-^2$ 
respectively. However due to the presence of $(n\cdot \partial)^{-1}$ in 
the original equation, the Hamiltonian itself 
depends on the energy eigenvalue $E$, which in the present case is equal 
to the mass of the particle. The solution for this case is given in
\cite{Dunn:2006}. The eigenvalues are found to be   
$E(\uparrow) = \sqrt{m_D^2+m_+^2-m_-^2}$ and 
$E(\downarrow) = \sqrt{m_D^2+m_+^2+m_-^2}$ for the positive energy (electron)
spinors with spin up and down respectively. Similar results are obtained
for anti-particles which are degenerate with particles. 
Here we use the non-relativistic limit and focus on the particle
states. 

The important point is that the energy eigenvalues
of the spin up and down states
are not degenerate. This means that the Hamiltonian is different for
these two states and hence does not really get diagonalized. In other
words the eigenvectors for spin up and down are eigenvectors of different
Hamiltonians and hence we are unable to construct a unitary operator which
will diagonalize the Hamiltonian.  
This means that the mathematical
structure of VSR is not consistent with the standard framework of quantum
mechanics. We are not sure how to mathematically solve this problem 
and do not pursue it further in full generality. 
However, as discussed below, we find that there are some limiting
 cases in which the problem is tractable.

The problem is obvious directly from Eq. \ref{eq:genDiracH}. 
We work in the Dirac-Pauli representation in which $\gamma^0$ is diagonal.
Due to presence of $E$ in the last two terms in this equation, diagonalization
of $H$ is possible only if (i) the operator $m_+^2\Gamma + m_-^2\Gamma\gamma^5$
is diagonal or (ii) All eigenvalues of H are degenerate. The first 
possibility is not realized for any choice of values of $m_+^2$ and $m_-^2$
while the second is found to be true if $m_-^2=0$. We see this directly
by the eigenvalues $E(\uparrow)$ and $E(\downarrow)$ given above. 
 Hence in this case the problem 
mentioned above no longer appears and the eigenvectors will correspond
to a unique Hamiltonian. We shall impose this condition 
for further analysis. In earlier work \cite{Dunn:2006} it has been argued that 
$m_-^2$ is very strongly constrained by observations. This may be 
correct but we have argued that it is really not possible to reliably
determine the experimental implications of the theory if $m_-^2\ne 0$. 
Hence it is not possible to impose reliable constraints on this parameter.       

We next consider the general case in which
 mass matrix $M_+^2$ is not diagonal. We continue to set $M_-^2=0$ based 
on the arguments presented above.
In this case the energy eigenvalues are clearly not degenerate since
different charged leptons have different masses. Hence the system can 
be solved only if the matrix $M_+^2$ is diagonal which is just 
the limit 
 discussed above. However if we assume that neutrino masses
are generated entirely by the VSR mass terms, then $M_+^2=U_e^\dagger M_\nu^2
U_e$  is necessarily
non-diagonal. Based on our arguments above, this case cannot be treated
reliably and we do not pursue it further. Only by assuming a diagonal form 
for the matrix $M_+^2$ can we impose reliable limits on the VSR parameters.
This of course significantly reduces the interest in further pursuing this 
formalism. Nevertheless we feel that it is an interesting theory
of Lorentz violation and continue to investigate its phenomenological
consequences. Furthermore 
it is possible that mathematical framework may be developed in future 
which may give a reliable solution to the problem in the general case.

The situation with VSR quark masses is similar. 
We have an effective Lorentz violating, VSR Lagrangian similar to 
Eq. \ref{eq:VSRmodel} with left and right handed terms both for
up and down type quark multiplets. This can be expressed as
\begin{eqnarray}
{\cal L}^q_{VSR} &=&  
{i\over 2}
\left(\begin{array}{cc}
\bar u_{i} & \bar d_{i}
\end{array}\right)_L 
\left[{\tilde M_{Lq}}^2\right]_{ij}
{\slashed{n}\over n\cdot D}
\left(\begin{array}{c}
u_{j}\\   d_{j}
\end{array}\right)_L\nonumber\\
&+&  {i\over 2} \bar u_{iR} \left[{\tilde M_{Ru}}^2\right]_{ij}
{\slashed{n}\over n\cdot D} u_{jR} \nonumber\\
&+&  {i\over 2} \bar d_{iR} \left[{\tilde M_{Rd}}^2\right]_{ij}
{\slashed{n}\over n\cdot D} d_{jR} 
\label{eq:VSRmodelq}
\end{eqnarray}
In principle, these 
mass matrices can be non-diagonal 
\cite{Dunn:2006,Alfaro:2015fha}.
Here we work in the basis in which the Dirac mass matrices are diagonal. 
For reasons discussed above for the case of leptons, the Hamiltonian in this
case also admits energy eigenvalues and eigenvectors only in the 
case in which the VSR mass matrices are diagonal. Hence we impose this
restriction for further analysis. Furthermore we set 
$M_-^2 = (M_L^2 - M_R^2)/2 = 0$ for reasons given earlier. 

\section{Limits based on torsion pendulum}
We next consider the limits that can be imposed on the VSR masses 
based on the spin pendulum experiment. Here we set $M_-^2=0$ and assume
that $M_+^2$ is diagonal both for 
quarks and leptons. The basic framework has 
been developed in \cite{Dunn:2006} which can be applied to electrons. 
In the non-relativisitic limit, the relevant term
in the effective Hamiltonian in the $n\cdot A =0$ gauge is given by 
\cite{Dunn:2006}
\begin{equation}
H_{VSR} = -\epsilon\mu_B  (\hat n\cdot\vec\sigma)
(\hat n\cdot\vec B)
\label{eq:HVSR}
\end{equation}
where $\vec B$ is the background magnetic field, $\hat n$ is the spatial
component of the vector $n^\mu$, $\mu_B= e\hbar/(2m_e c)$,  
$\epsilon = m_+^2/m_e^2$, $m_e$ is the electron mass and $m_+^2$ is
the VSR contribution to the electron mass. 
By using the results of Penning trap experiment with a single trapped electron 
\cite{Mittleman:1999} it was found that
 $\epsilon \lesssim 10^{-11}$ \cite{Dunn:2006}. It may be possible
to impose a more
stringent limit by using the experimental results
on torsion pendulum \cite{Heckel:2006}. However it is not possible to
directly use the limits given in \cite{Heckel:2006}. This is because those
limits have been obtained by assuming that the effect has a time
period of 1 sidereal day. However Eq. \ref{eq:HVSR} shows that the effect
is more complicated. In particular, as discussed below,
 it shows two oscillations in 1 
sidereal day.  

Let us denote the equatorial coordinate system by $xyz$ and the laboratory
system by $abc$. We choose coordinates such that z-axis is parallel to the 
rotation axis of the Earth and x-axis points towards the vernal equinox. 
Let the equatorial coordinates of the vector $\hat n$ be  $(\theta_e,\phi_e)$  
where $\phi_e$ is the Right Ascension and $\theta_e$ is the polar angle 
(Declination = $\pi/2 - \theta_e$). 
Let the unit vectors along the
axis $a,b,c$ of the local frame be $\hat i, \hat j$ and $\hat k$ respectively.  We take the vector $\hat k$ to point vertically upwards and vectors
$\hat i$ and  $\hat j$ tangential to the surface pointing towards north and
west respectively.  
The two coordinate systems are related by the formula
\begin{eqnarray}
\hat x &=& \cos(\theta+\alpha)\hat j-\sin(\theta+\alpha)(\cos\lambda\hat k
-\sin\lambda\hat i)\nonumber\\
\hat y &=& \sin(\theta+\alpha)\hat j+\cos(\theta+\alpha)(\cos\lambda\hat k
-\sin\lambda\hat i)\nonumber\\
\hat z &=& \cos\lambda\hat i + \sin\lambda \hat k
\end{eqnarray}
where $\theta=\Omega t$, $\alpha$ is the Right Ascension of $\hat j$ at 
$t=0$, $\lambda$ is the latitude of the observer, $\Omega=2\pi/T_0$ and
$T_0$ is one sidereal day.

The spin of the torsion pendulum used in  \cite{Heckel:2006} is aligned
horizontally. Hence we set  
the magnetic moment $\vec m$ of the pendulum equal to $m_1\hat i + m_2\hat j$.  
The magnetic field $\vec B = B_1\hat i + B_2\hat j$ also points in the same direction as $\vec m$
and hence $B_1/B_2 = m_1/m_2$. 
Using this we can determine the torque experienced by a single electron 
due to VSR effects. 
The torque about the local normal is found to be
\begin{eqnarray}
\tau_b &=& \epsilon\hat n\cdot \vec B\bigg\lbrace m_1\sin\theta_e
\cos(\theta+\alpha-\phi_e)\nonumber\\ 
&-& m_2\sin\theta_e\sin\lambda\sin(\theta+\alpha-\phi_e) \nonumber\\
&-& 
m_2\cos\theta_e\cos\lambda
\bigg\rbrace 
\label{eq:torque}
\end{eqnarray}
where 
\begin{eqnarray}
\hat n\cdot\vec B &=& B_1\bigg[
\sin\lambda\sin\theta_e
\sin(\theta+\alpha-\phi_e) + \cos\lambda\nonumber\\
&\times &\cos\theta_e 
\bigg] + B_2 \sin\theta_e\cos(\theta+\alpha-\phi_e)
\end{eqnarray}

\begin{figure}[h]  
\begin{center}  
\includegraphics[width=0.45\textwidth]{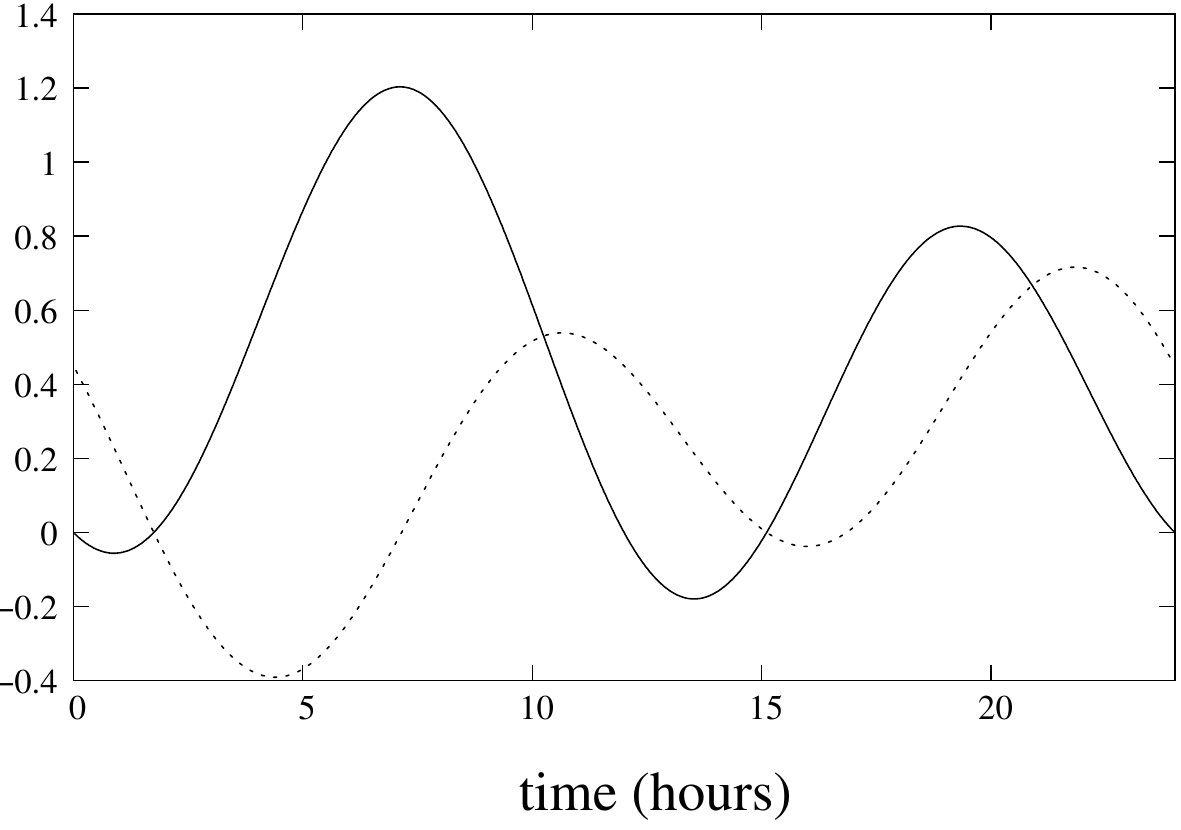}
\caption{\small \sl The predicted signal for the torques generated
in the torsion pendulum \cite{Heckel:2006} in arbitrary units for
a randomly chosen set of parameters $\theta_e=0.6\pi$ and 
$\alpha -\phi_e = \pi/2$. 
The solid and dotted curves refer to $\beta_W$ and $\beta_N$ using 
the notation of \cite{Heckel:2006}.  
The latitude $\lambda$ has been set equal to that of the observer location.
We have also set $m_1/m_2 = B_1/B_2 = 1$.   }  
\label{fig:torsion1}  
\end{center}
\end{figure}

The experimentalists \cite{Heckel:2006} 
split data into torques generated by the north
and west components of the effective field $\vec\beta$ 
which couples to $\vec m$, such that the energy $E\propto -\vec\beta\cdot \vec m$. 
In our case this corresponds to $n_1$ and $n_2$ respectively times an 
overall factor $\epsilon \hat n\cdot \vec B$. The
corresponding torques generated by these components are given by
the terms proportional to $m_2$ and $m_1$ respectively in Eq. \ref{eq:torque}.
The overall expression is different from 
 what is assumed in the analysis performed in \cite{Heckel:2006}. 
Hence we suggest that the data should be reanalyzed in order to 
impose constraints on VSR parameters. In Fig. \ref{fig:torsion1} 
we show a representative
graph of our results setting $m_1/m_2=1$.  
We clearly see that the signal shows two oscillations within one sidereal
day. Hence a dedicated analysis is needed in order to obtain reliable
limits on VSR parameters. Assuming that no signal is found, the torsion
pendulum experiment \cite{Heckel:2006} will impose limit on $\epsilon$
such that $\epsilon \lesssim 10^{-20}\ {\rm eV}/(\mu_BB)$ which 
will imply $\epsilon \lesssim 10^{-15}$.    

\subsection{Limits on VSR nucleon mass using clock comparison experiments}

We next consider bounds that can be imposed on VSR contribution
to nucleon mass using clock
comparison experiments \cite{Hughes:1960,Prestage:1985,Berglund:1995,Kostelecky:1999mr} with polarized nucleons \cite{Bear:2000,Phillips:2001,Cane:2004}.   
Let $\tilde m_n^2$ be the VSR contributions to the
nucleon mass. We assume isospin symmetry and set the proton and neutron mass
equal to one another. Furthermore we treat nucleon as a Dirac particle
with an effective magnetic moment described by the g-factor of proton
or neutron. We also ignore nuclear effects which have to be included for 
a detailed fit. As argued earlier we only allow contributions which
are proportional to $\slashed {n}$ and set the contribution proportional to
$\slashed {n} \gamma_5$ to zero. The VSR nucleon mass may be
 related to the up and
down quark VSR masses, denoted by $\tilde m_q^2$, 
of up and down quarks by a form factor $G_n$. 
Hence we expect $\tilde m_n^2    
= G_n \tilde m_q^2$. 

\begin{figure}[h]  
\begin{center}  
\includegraphics[width=0.45\textwidth]{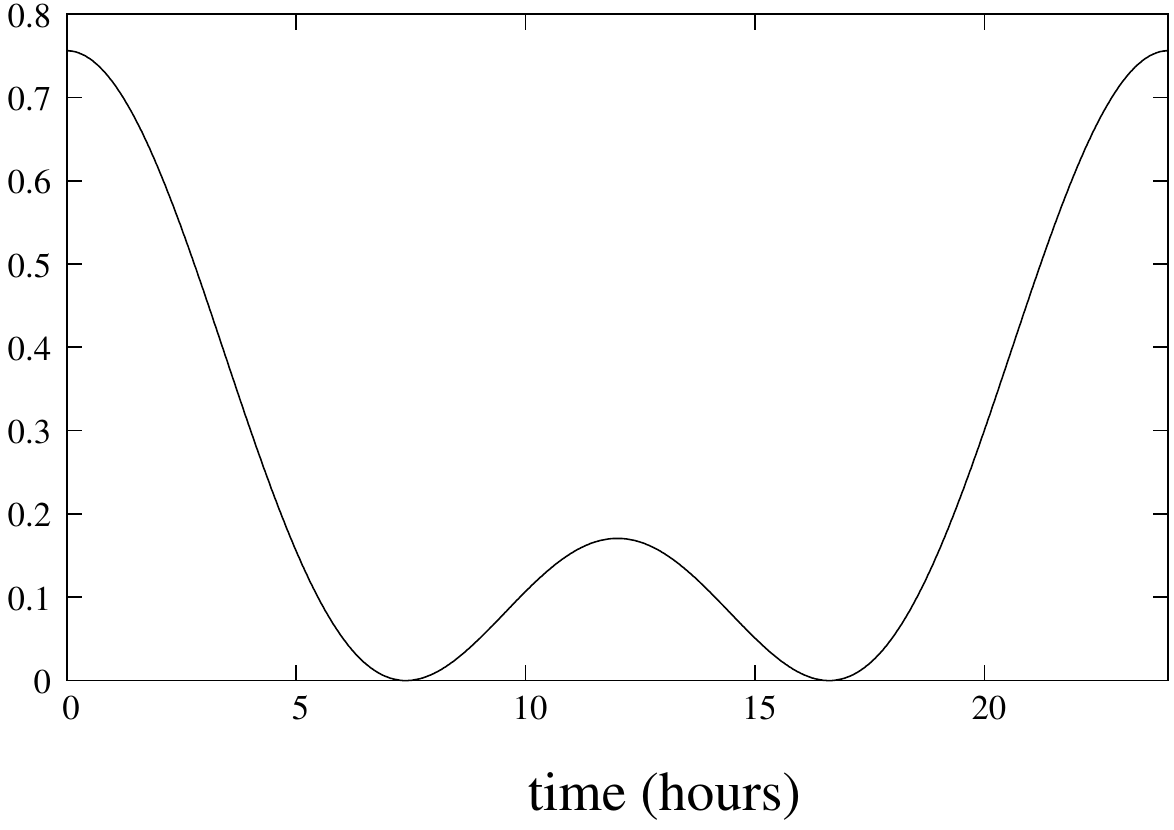}
\caption{\small \sl The predicted signal for $\xi$, the shift in the
precession frequency of nucleon (in arbitrary units) due to VSR 
effects. Here we have chosen the same set of parameters
$(\theta_e, \alpha -\phi_e, \lambda)$ as in
Fig. \ref{fig:torsion1}.  
  }  
\label{fig:deltaom}  
\end{center}
\end{figure}

We consider an experiment with polarized 
neutrons or protons, with their spins vertically upwards.  
The VSR effect will lead to a shift in the precession frequency of
the nucleons. The effect has already been considered in \cite{Dunn:2006}
for the case of electrons. Essentially we can incorporate the 
effect by defining an  effective magnetic
field $\vec B' = \vec B + (2/g)\epsilon_n \hat n (\hat n\cdot\vec B)$,  
where $g$ is the g-factor of the particle, proton or neutron,
$\epsilon_n=\tilde m_n^2/m_n^2$ and $m_n$ is the nucleon mass.
The magnetic field in this case points along $\hat k$ and hence
 the frequency gets shifted by the factor $(1+\xi)$
where $\xi = (2/g)\epsilon_n (\hat n\cdot \hat k)^2$. We   
obtain
\begin{equation}
\hat n\cdot \hat k = \cos\theta_e\sin\lambda - 
\sin\theta_e\cos\lambda \sin(\theta+\alpha-\phi_e) 
\end{equation}
We plot $\xi$ for arbitrarily chosen parameters in Fig. \ref{fig:deltaom}. 
We again clearly see that the signal is not just a simple
sinusoidal variation with a period of a sidereal day. Instead
we see two oscillations with varying amplitude within one sidereal day.
Hence a dedicated search is needed in order to constrain the VSR parameters.
Assuming that the signal is absent in the data, we can obtain the bound
$\epsilon_n\mu_N B\lesssim 10^{-31}$ GeV using the experimental data 
from \cite{Cane:2004}, where $\mu_N = e\hbar/(2m_n c)$. This will lead to the
limit $\epsilon_n\lesssim 10^{-11}$ where we have used $B\approx 1.5$ G. 
This leads to $\tilde m_n^2\lesssim 10^7$ eV$^2$.

\section{Pion Decay}

We next consider implications of VSR for elementary particle physics 
experiments. Here we are primarily interested in effects which arise 
due to rotation of Earth. We shall illustrate the effect by considering
the decay of pion as an example. Similar effects are expected in other
processes. 
The decay amplitude within the
SM can be computed by introducing the following
effective interaction term:
\begin{eqnarray}
{\cal L}_{\pi,SM}= V_{ud}\frac{G_{f}}{\sqrt{2}}{f}_{\pi}\partial_{\mu} \pi^- \bar{\psi_l}\gamma^{\mu}(1-\gamma^5)\psi_\nu + h.c. \,, 
\label{eq:Lag_SM}
\end{eqnarray}
where $\pi^-$, $\psi_l$ and $\psi_\nu$ represent the charged pion, 
charged lepton and the neutrino fields respectively. Here $f_\pi = 132$ MeV
is the pion decay constant, $V_{ud}=\cos\theta_c$ is the CKM matrix element
and $\theta_c$ the Cabibbo angle.   
 This leads to the standard formula
for the weak differential decay rate of pions.

The basic Lorentz violating, VSR invariant terms for quarks are given in 
Eq. \ref{eq:VSRmodelq}. Due
to the presence of gauge covariant derivative, the VSR terms also lead to 
Lorentz violating VSR invariant interaction terms of fermions with 
electroweak gauge bosons as well as gluons for the case of quarks
 \cite{Cheon:2009zx,Alfaro:2015fha}. 
As argued earlier we require the resulting VSR quark mass matrices to
be diagonal and furthermore set $M_-^2=0$. 
As we shall show below VSR terms
in Eq. \ref{eq:VSRmodelq}
  lead to an  
  effective Lagrangian density for the coupling of pions with
leptons which   
 can be written as 
\begin{eqnarray}
{\cal L}_{\pi}&=& {\cal L}_{\pi,SM} + \tilde g \left(\frac{n_{\mu}}{n.\partial}\pi^- \right)\bar{\psi}_l \gamma^{\mu}(1-\gamma^5)\psi_\nu \nonumber\\ 
&+& h.c. ,
\label{eq:Lag}
\end{eqnarray}
where $n^{\mu}$ is given by Eq. \ref{eq:vectorn}.  
The first term gives the standard decay amplitude for pion. The second term 
respects SIM(2) but violates Lorentz invariance due to the 
 presence of preferred axis $n^{\mu}$ \cite{Cohen:2006ir}. 

\begin{figure}[h]  
\begin{center}  
\includegraphics[width=0.50\textwidth]{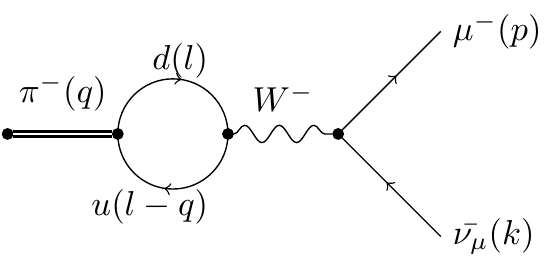}
\caption{\small \sl The Lorentz violating matrix term arises from quark VSR modified propagator and lepton-W boson vertex correction in the VSR modified SM Lagrangian. }  
\label{fig:piondecay}  
\end{center}
\end{figure}

We next provide a justification for the Lorentz violating effective operator 
in Eq. \ref{eq:Lag}
using linear sigma model for strong interactions and VSR modified 
SM Lagrangian which has mass terms of the form given in 
Eqs. \ref{eq:VSRmodel} and \ref{eq:VSRmodelq} for all fermions.  
Alternatively we may use a model pion wave function \cite{Jain:1993qh} for
this calculation. Here we are primarily
interested in demonstrating that this loop leads to a non-zero answer and hence
our use of linear sigma model is justified. However a quantitatively
 reliable estimate of
this loop is not possible due to the standard uncertainties in handling
strong interactions. 
We consider linear sigma model at the quark level which contains the fermion 
field multiplet $\begin{pmatrix} u  \\ d \end{pmatrix} $ and the meson
multiplet $\sigma+i\tau\cdot \pi\gamma^5$, where $\sigma$ is a scalar
field. The model leads to an
interaction between pseudoscalar pion field and the quark doublet of
the form
\begin{equation}
{\cal L}_\sigma= g^{\prime}\begin{pmatrix} \bar{u}_i  & \bar{d}_i\end{pmatrix}
(\sigma+i\tau.\pi \gamma^5) \begin{pmatrix} u_i  \\ d_i \end{pmatrix}, 
\end{equation}
where $g^{\prime}$ is the coupling and $i$ is the color index. 
The pion decay into a lepton pair
can be represented by the diagram shown in Fig. \ref{fig:piondecay}.
The coupling of pion with the up and down quarks is given by the 
linear sigma model. 
The Lorentz violating VSR invariant contributions arise due to 
the modification to the up and down quark propagators and the 
interaction vertex of $W$ boson with quarks. 
The VSR modified fermion propagator can be written as, 
$$\frac{\slashed{l}+m-\frac{\tilde m^2}{2}\frac{\slashed{n}}{n.l}}{l^2-m^2-
\tilde m^2+i\epsilon}\,.$$  
where $m$ is the standard mass arising due to a Lorentz invariant term
and $\tilde m$ is the VSR mass.  
We shall assume that for all fermions, except the neutrinos, $\tilde m<< m$.
The interaction terms arise due to the gauge covariant derivative in
 Eq. \ref{eq:VSRmodelq}. We expand 
$1/n\cdot D$ in powers of the gauge coupling and keep only the leading order 
term in this coupling.  
The modified lepton-W boson vertex is found to be 
$$\frac{g}{\sqrt{2}}\left[\gamma^{\mu}+\tilde m^2 \frac{\slashed{n}n^{\mu}}
{2 (n\cdot l) (n\cdot (l-q))}\right] (1-\gamma^5)\,. $$

The Feynman amplitude shown in Fig. \ref{fig:piondecay}
generates an effective vertex between the pion and the leptons.
We can obtain the effective coupling by 
evaluating the Feynman amplitude for the quark loop in Fig. \ref{fig:piondecay},
which can be expressed as,
\begin{widetext}
\begin{eqnarray}
g^{\prime}\frac{g}{\sqrt{2}}\int\frac{d^4l}{(2\pi)^4} tr \Big[\frac{\slashed{l}+m_d-\frac{\tilde m_{q}^2}{2}\frac{\slashed{n}}{n.l}}{l^2-(m_d^2+\tilde m
_{q}^2)+i\epsilon}\left(\gamma^{\mu}+\tilde m_{q}^2 \frac{\slashed{n}n^{\mu}}{2 n.l n.(l-q)}\right) (1-\gamma^5) \frac{(\slashed{l}-\slashed{q})+m_u-
\frac{\tilde m_{q}^2}{2}\frac{\slashed{n}}{n.(l-q)}}{(l-q)^2-
(m_u^2+\tilde m_{q}^2)+i\epsilon}\gamma^5 \Big]
\end{eqnarray} 
\end{widetext}
where $\tilde m_q=\tilde m_u=\tilde m_d$. We note that for the left
handed up and down quarks the VSR masses have to be equal by gauge invariance.
In evaluating this loop 
 we consider terms only up to order $\tilde m_{q}^2$ because of our
assumption $\tilde m_q<< m_u, m_d$.  
It is also important to notice that the VSR invariant, Lorentz violating 
terms are
nonlocal.

The Lorentz violating or nonlocal part of the above expression  becomes
\begin{widetext}
\begin{eqnarray}
&&g^{\prime}\frac{g}{\sqrt{2}}\tilde m_{q}^2n^{\mu}\int \frac{d^4l}{(2\pi)^4}  \Big( \frac{4 m_d \, n\cdot q }{n.l \hspace{3pt} n.(l-q) (l^2-m_d^2+i \epsilon)((l-q)^2-m_u^2+i\epsilon)} \nonumber \\
&& +    \frac{4 m_u }{n.(l-q) (l^2-m_d^2+i \epsilon)((l-q)^2-m_u^2+i\epsilon)}    -\frac{4 m_u }{n.l (l^2-m_d^2+i \epsilon)((l-q)^2-m_u^2+i\epsilon)}\Big)
\end{eqnarray}
\end{widetext}
After integrating over $l$ the result can only depend on the momentum of the
pion, i.e. $q$. Hence the integral leads to an overall factor of $1/n\cdot q$.
This gives us an effective interaction of the form given in Eq. 
\ref{eq:Lag}. We next perform this integral in the rest frame of pion.  

The presence of nonlocal term in the fermion propagator inside the loop results in infrared (IR) divergences. It is not practical to use Feynman parametrization for evaluating the above integral because of the presence of nonlocal terms, i.e. $\frac{1}{n.l}$ and $\frac{1}{n.(l-q)}$. 
There doesn't exist any reliable procedure for handling these infrared
divergences in the literature. 
We handle this divegence 
by adding a small imaginary part to the mass of
the particle. This amounts to adding a small imaginary part to the energy
of an on-shell particle. 
Hence we replace 
\begin{equation}
{1\over n\cdot l} \rightarrow {1\over n\cdot (l+i\epsilon')}
\end{equation}
where $\epsilon'$ is a four vector with time component ($\epsilon'_0>0$) 
non-zero and 
space components zero. 
This prescription may be justified by considering the action of the
operator $1/n\cdot \partial$ on a charged scalar
field $\phi$. The Fourier decomposition of this may be expressed as
\begin{equation}
\phi = \int {d^3k\over (2\pi)^32\omega_k}\left[ e^{-ik\cdot x} a(k) + 
e^{ik\cdot x} b^\dagger(k)\right]
\end{equation} 
Action of the operator $1/n\cdot \partial$ on this field should result
in the factor $1/n\cdot k$. In order to explicitly implement this we
need a prescription for $1/n\cdot \partial$ which essentially is 
equivalent to an integral. We follow a prescription
which is analogous to the one used in Ref. \cite{2012PhRvD..85j5009V}. 
For the positive frequency
part, we set 
\begin{equation}
{1\over n\cdot \partial}f^+(x) \rightarrow \int_{-\infty}^{x_+} dx'_+ f^+(x')
\end{equation} 
whereas for the negative frequency we use 
\begin{equation}
{1\over n\cdot \partial}f^-(x) \rightarrow -\int_{x_+}^{\infty} dx'_+ f^-(x')\,.
\end{equation} 
Here $x_+ = (t+z)/2$. Let us now apply this operator to the positive 
frequency part. We obtain
\begin{widetext}
\begin{equation}
{1\over n\cdot \partial}\phi^+(x) =
\int {d^3k\over (2\pi)^32\omega_k} 
{1\over -ik_-}\left[e^{-ik_-x_+} - e^{ik_-\infty}\right]e^{-ik\cdot x+ik_-x_+}
a(k)
\end{equation} 
\end{widetext}
where $k_- = k^0-k^3 = n\cdot k$. This clearly gives us the expected result
as long as the second term in the bracket goes to zero. This is true if
$k^0$ contains a small imaginary part as prescribed above. Similarly
we can check that the negative frequency part leads to the expected
result with our prescription. 
  
The loop integral is evaluated by performing the integral over $l^0$
analytically and over the spatial components numerically. The
numerical calculations are performed by using a small non-zero value
of the infrared regulator $\epsilon'_0$ of the order of a few MeV. We have
verified that the final result is insensitive to the precise choice
of this regulator provided it is sufficiently small. 
 For $m_d=4.8 $ MeV,  $m_u=2.3 $ MeV and $m_\pi =139.5 $ MeV, we obtain the result
 $-N_cg^{\prime}\frac{g}{\sqrt{2}}\tilde m_{q}^2n^{\mu}\times 0.092$ 
where $N_c=3$ is the number of colors. 
Hence the loop gives a non-zero result and generates an effective vertex
given in Eq. \ref{eq:Lag}. As already mentioned we can only trust this 
result qualitatively and not quantitatively due to our inability to
reliably handle strong interactions.

\subsection{pion decay in the rest frame}

In this section we compute the decay rate of charged pion  
($\pi^-(q) \rightarrow \mu^-(p) + \bar{\nu}_{\mu}(k)$) in its rest frame 
within the VSR framework. We consider $|\mathfrak{M}|^2$ upto leading order in $\tilde{g}$ because the
 LV parameters are expected to be very small. We obtain 
 \begin{widetext}
 \begin{eqnarray}
|\mathfrak{M}|^2=4G_{f}^2\mathfrak{f}_{\pi}^2V_{ud}^2m_{\mu}^2(p.k) + \frac{16}{\sqrt{2}}\tilde{g}G_{f} f_{\pi}V_{ud}\left( \frac{(n.p)(q.k)}{n.q} + \frac{(n.k)(p.q)}{n.q}-\frac{(n.q)(p.k)}{n.q}\right) + O(\tilde{g}^2)
\label{eq:amp1}
\end{eqnarray}
 \end{widetext}
This is valid in general and not just the rest frame.

As explained above, we can 
always make a SIM(2) transformation to the rest frame of
a particle. Our action is invariant under this transformation, although 
the vector $n^\mu$ changes by an overall constant. 
However the change cancels out in the amplitude.
Here we work in a frame ($S$) in which 
the vector $n^\mu$ is given by Eq. \ref{eq:vectorn} up to an 
overall constant.   
The LV contribution is assumed to arise
entirely from the interaction term in Eq. \ref{eq:Lag}.
We point out that the VSR invariant quadratic terms do not change the
dispersion relations \cite{Cohen:2006ir}. Hence the kinematics of the
incoming and outgoing particles remain unchanged. The dominant LV contribution
to the differential decay rate arises due to SM and LV interference 
term. This leads to a contribution proportional
to $\tilde g(1+\cos\theta)$ where $\theta$ is the angle between the  
muon three momentum and the $z$-axis
in the fundamental frame $S$.

We next impose a direct 
limit on $\tilde g$ by assuming that the standard observed 
value of the pion decay rate arises entirely from the Standard Model
and demanding that the LV terms give a contribution less than the error
in the observed value. A more detailed limit by studying the angular
distributions of the final state can also be imposed. In the next section
we shall work out the theoretical formalism required for such a study. 
However a detailed implementation can only be performed by an experimental
group and is beyond the scope of current paper.
  We point out that here we are only interested in direct experimental
limit that can be imposed on this LV parameter. Through loop corrections
this parameter may lead to LV contribution to electron propagator. However
such contributions only add to the LV parameters in the leptonic sector
of the action which can be adjusted to agree with experimental limits.   
This might require some fine tuning of parameters. 

The life time ($\tau$) of the charged pion is $(2.6033\pm 0.0005)\times 10 ^{-8} $s
\cite{Patrignani:2016xqp}. The uncertainty in the  
theoretical calculation of the pion decay rate is approximately 0.2\%.
Hence we see that the theoretical uncertainty dominates.  
This leads to the limit, $\tilde g < g_0$ where $g_0= 2.2\times 10^{-11}$ GeV.
We can relate this to the VSR up and down quark mass parameter $\tilde m_q$  
through the loop shown in Fig. \ref{fig:piondecay} and impose a limit
on this parameter. 
We fix the linear sigma model parameter $g'$ by using the standard relation
$M_q=g'f_\pi$, where $M_q\approx 330$ MeV 
\cite{Capstick:2000qj} is the constituent quark mass of up and 
down quarks.
We obtain $\tilde m_{u,d}\lesssim 3$ MeV. We notice that the limit
is not very stringent and is much weaker than that obtained by using
clock comparison experiments assuming that the nucleon form factor $G_n$
is of order unity.   

The relative change of differential decay rate
 can be expressed as 
 \begin{eqnarray}
 \Delta&=&\frac{\frac{d\Gamma}{d\Omega}|_{\tilde g\not{=}0}-\frac{d\Gamma}{d\Omega}|_{\tilde g=0}}{\frac{d\Gamma}{d\Omega}|_{\tilde g=0}}\nonumber\\ 
 &=& {2\sqrt{2}\tilde g\over f_\pi m_\pi^2 G_f|V_{ud}|} \left[1  +\cos\theta\right]\,.
\label{eq:Delta1}
 \end{eqnarray}
Hence we find  
 that even in the rest frame of pion, the muon distribution is not 
isotropic and 
 depends on the polar angle $\theta$ due to the LV contributions.
 The dependence provides a qualitatively new test 
of LV theories which respect VSR. 

\subsection{Kaon Decay}

The above formalism can be directly applied to the charged kaon decay
$K^-(q)\rightarrow \mu^-(p) +\bar{\nu}_\mu(k)$. We compute the loop integral
corresponding to the quark loop shown in Fig. \ref{fig:piondecay} by
replacing the down quark with a strange quark. The Lorentz violating
part of the loop integral in this case is found to 
be $ N_c g^{\prime} \frac{g}{\sqrt{2}} \tilde {m}_{q}^2\times 2.32 $.  
Furthermore we use $M_s=g'f_K$ where $M_s\approx 550$ MeV 
\cite{Capstick:2000qj} 
is the constituent mass of the strange quark. Using the known 
uncertainty in the kaon decay rate we impose the
limit $\tilde g < 1.0\times 10^{-10}$ GeV. In this case the theoretical
and experimental errors are comparable to one another and we add the
two in quadratures in order to obtain the limit on $\tilde g$. 
This leads to a limit
$\tilde m_s\lesssim 1.3$ MeV on the VSR contribution to strange quark mass.

  
\subsection {pion decay in the laboratory frame}
\begin{figure}
  \centering
    \includegraphics[width=7.30cm]{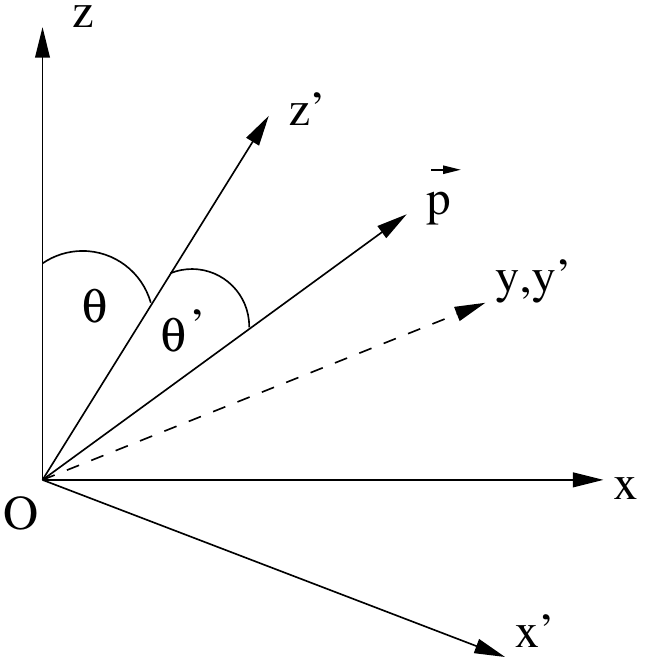}
    \caption{Here $z$ denotes the preferred axis and $x,y$ some chosen 
coordinate axes. The beam direction is taken to be along the $z'$-axis which
makes an angle $\theta$ relative to $z$-axis. The axis $x'$ is chosen
to lie in the $x-z$ plane. Hence the $y$ and $y'$ axes, pointing into the
plane of the paper, coincide with one another. The momentum of the
muon, denoted by $\vec p$, makes an angle $\theta'$ relative to the
$z'$-axis.
}
    \label{fig:vsr-axis}
  \end{figure}

 In this section, 
we determine the differential decay rate assuming that pion has non-zero
momentum in the laboratory frame. 
It is useful to define two frame $S$ and $S'$, both at rest with respect to
the laboratory.
In frame $S$, $n^\mu$ is
given by Eq. \ref{eq:vectorn} up to an irrelevant overall constant. 
Let us now consider a beam of pions moving along the $z^{\prime}$ direction making an angle $\theta$ with the preferred axis, as shown in Fig. \ref{fig:vsr-axis}. Let $q$, $p$ and $k$ denote the momenta of $\pi^-$,
$\mu^-$ and  $\bar{\nu}$ respectively. Here $xyz$
and $x'y'z'$ refer to $S$ and $S'$ 
coordinate systems 
respectively.   
We have used rotational symmetry about the $z$-axis in 
the frame $S$ in order
to choose $x$-axis such that 
$y'$ is aligned with the $y$-axis. Hence $z'$ and $x'$ lie in the $x-z$ plane. 
The final state muon makes an angle $\theta^{\prime}$ w.r.t. the beam, i.e. the  $z^{\prime}$ axis. 

\begin{figure}
  \centering
    \includegraphics[width=7.30cm]{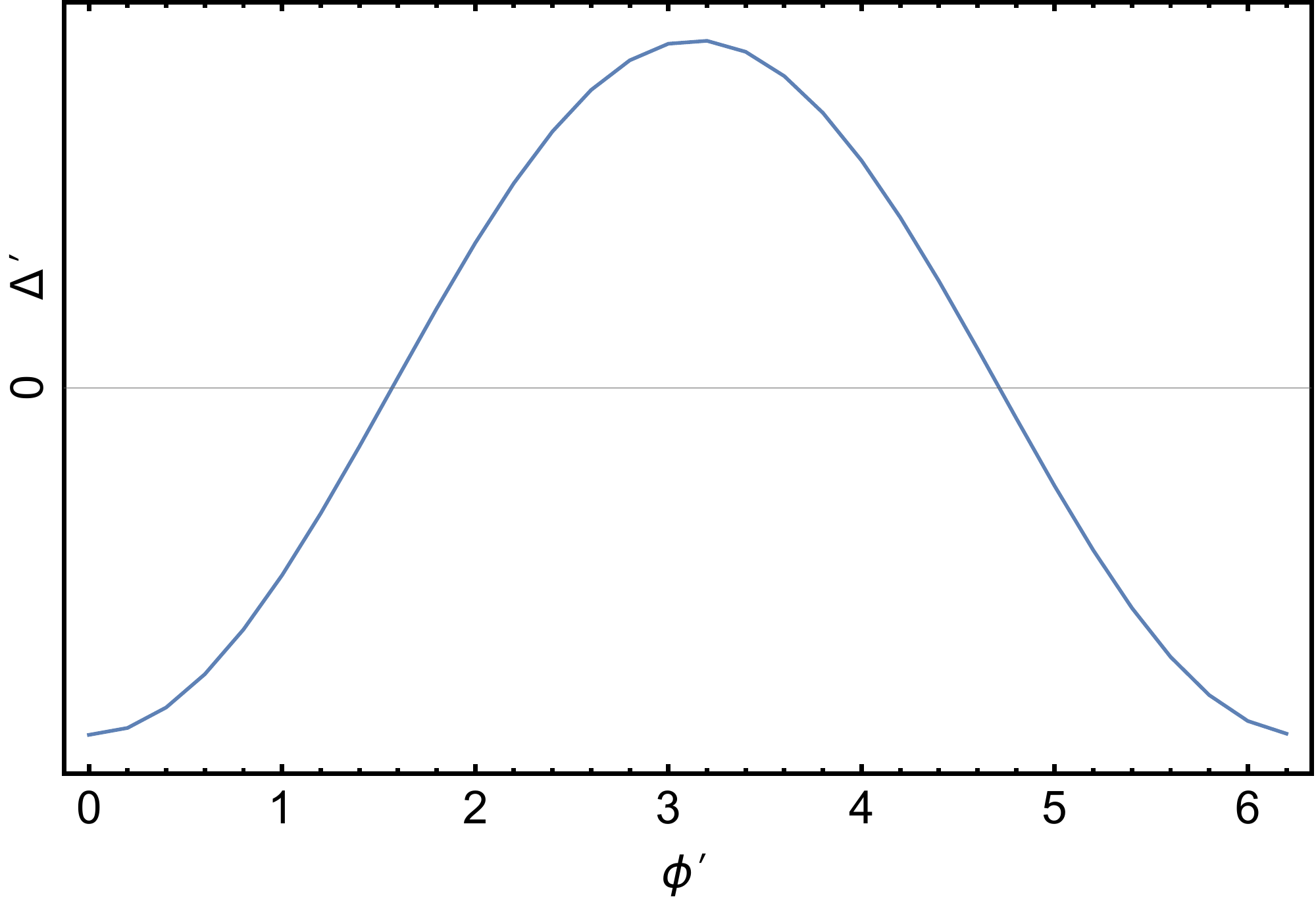}
    \caption{The azimuthal angle $\phi'$ dependence of the final state
muon distribution. The observable $\Delta^{\prime}$ is defined in 
Eq. \ref{eq:Deltaprime}. The amplitude of the effect depends on the 
unknown VSR parameter $\tilde g$ or equivalently $\tilde m_{u,d}^2$ for which we only have an upper limit.
}
    \label{fig:Figd}
  \end{figure}

We find that the differential decay rate 
picks up a small correction to the $\theta'$ dependence of
the decay rate due to the LV term. Furthermore it induces a $\phi'$
dependence of the final state muon distribution, which is absent in 
the SM. 
The $\phi'$ dependence of the decay rate can be quantified by defining
\begin{eqnarray}
\Delta^{\prime}=\frac{\frac{d\Gamma}{d\phi^{\prime}}-\Gamma_{avg}}{\Gamma_{avg}}
\label{eq:Deltaprime}
\end{eqnarray}
where 
$\Gamma_{avg}=\frac{1}{2 \pi}\int_0^{2\pi}\frac{d\Gamma}{d\phi^{\prime}}d\phi'$
is the decay rate averaged over $\phi'$. 
In Fig.\eqref{fig:Figd} we plot $\Delta'$ as a function of $\phi'$
for the choice of parameters, pion energy 
  $E=200$ MeV and $\theta=\pi/4$. 
We see that the distribution peaks at $\phi'=\pi$ and is minimum at
$\phi'=0$. From Fig. \ref{fig:vsr-axis} we see our choice of coordinate
system is such that the beam axis, i.e. $z'$, lies in the $x'-z$ plane. 
Hence $\phi'$ is the azimuthal angle in the $x',y',z'$ 
coordinate system which is chosen such that $z'$ lies
in the $x'-z$ plane

\subsection{Daily Variation} 
The angle between the preferred axis and the beam direction is expected to 
change with time due to rotation of Earth. Due to this change the 
contribution to the differential decay rate arising from the LV term 
is expected to show periodic variation with a period of 1 sidereal day.  
Both the observables $\Delta$ and $\Delta'$ are expected to show
time dependence. In particular we expect that the peak position of
$\Delta'$ as a function of the azimuthal angle $\phi'$ in laboratory
frame will show a periodic shift with time.

Let us assume that an observer is
located at the latitude $\lambda$. We choose a local laboratory 
coordinate system at this location, denoted by $x''y''z''$. Here $z''$ is
along the direction of the beam and $y''$ is chosen along the local 
vertical. It is also convenient to define another local frame $x'y'z'$
such that $z'$ is along the beam direction, i.e. same as $z''$ and 
$x'$ lies in the $z-z'$ plane. We denote the angle between $z$ and $z'$
by $\theta$ as shown in Fig. \ref{fig:vsr-axis}. Hence $\hat z\cdot\hat z'
=\hat z\cdot\hat z'' = \cos\theta$.  The $x'$-axis lies in the same plane
as $z$ and $z'$ (or $z''$). The $x'-y'$ and $x''-y''$
planes coincide and we denote the angle between $x$ and $x'$ as $\beta$. 
Using this we obtain
\begin{equation}
\hat z 
       = \cos\theta\hat z'' - \sin\theta\left(\cos\beta\hat x'' +\sin\beta
\hat y''\right)
\end{equation}
The coordinates $x'y'z'$
at any particular time are exactly the same as in Fig. \ref{fig:vsr-axis}. 
Hence once we obtain the angle $\theta$, which is time dependent, we can
obtain the differential decay rate in this frame at any particular
time using the formulism described earlier. In
this frame the peak in the $\phi'$ distribution occurs at $\phi'=\pi$
as shown in Fig. \ref{fig:Figd}. We next need
to transform to the laboratory frame $x''y''z''$. This simply amounts to a
rotation about the $z'$ (or $z''$) axis by an angle 
$-\beta$. Hence in this frame the peak occurs at $\phi''=\pi-\beta$. 

\begin{figure}
  \centering
    \includegraphics[width=7.30cm]{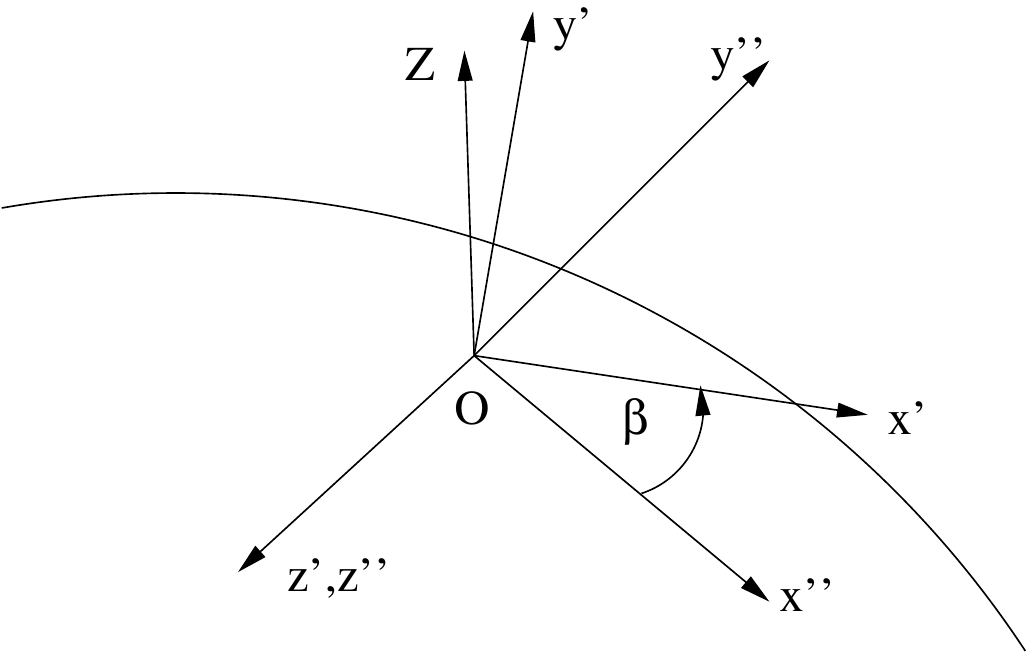}
    \caption{The laboratory coordinates $x''y''z''$ and the local coordinates
$x'y'z'$ at the position of the observer $O$ located at latitude $\lambda$. 
Here $z'$ and $z''$ are along the beam direction. The rotation axis of 
Earth $Z$ is also shown.  
 The $y''$ coordinate is
taken to be the local normal, pointing upwards. The $x'$ direction is chosen
such that it lies in the $z-z'$ frame as shown Fig. \ref{fig:vsr-axis}.}
    \label{fig:localframe}
  \end{figure}

\begin{figure}
  \centering
    \includegraphics[width=7.30cm]{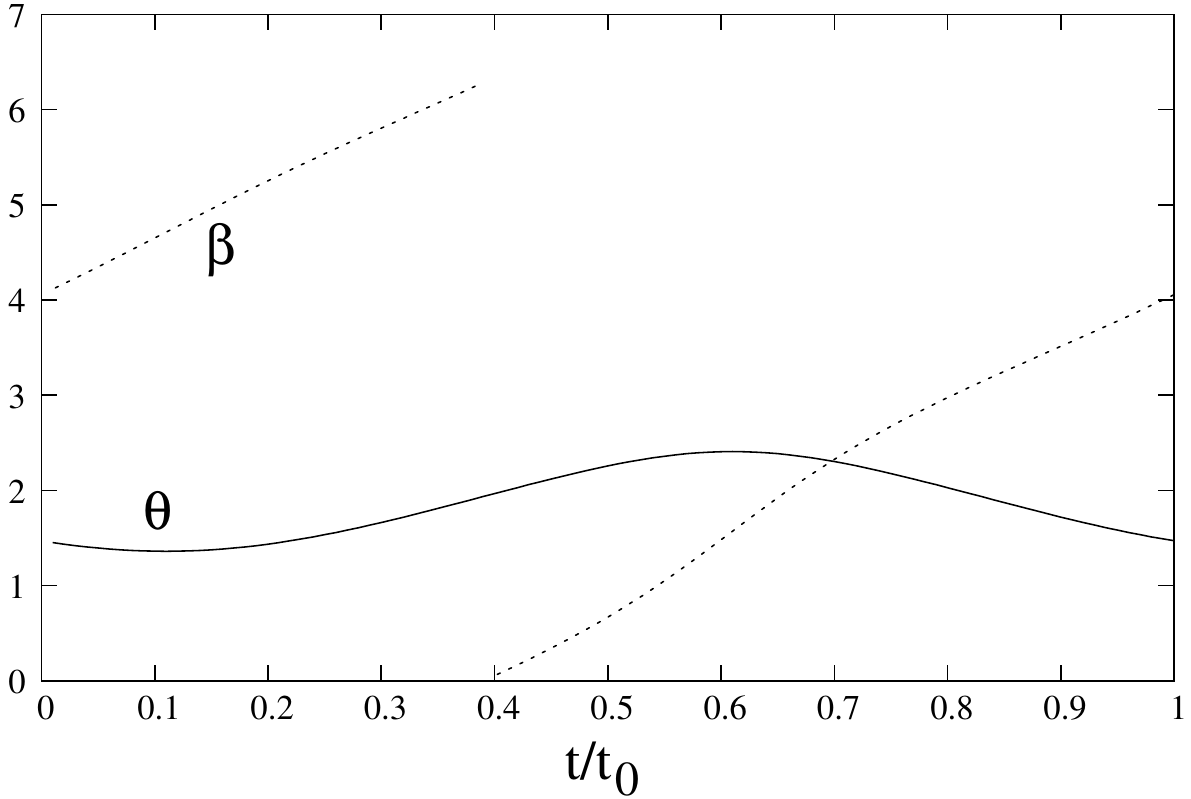}
    \caption{The time dependence of $\theta$ (solid curve) 
and $\beta$ (dotted curve), $0\le \beta<2\pi$, as a function of time. 
Here $t_0$ is equal to 1 sidereal day. The observer is located at $\lambda=
30^o$ and the remaining angles (in radians) are chosen as, $\theta_p=0.4\pi$,
$\phi_p=0.3\pi$ and $\alpha=0.1\pi$. The peak position in the $\phi''$
distribution occurs at $\pi-\beta$.   }
    \label{fig:tdep}
  \end{figure}

We next determine the time dependence of the angles $\theta$ and
$\beta$ due to the rotation of Earth. We use the astronomical 
equatorial system as our fixed coordinate system denoted by $XYZ$. 
In this case the $Z$-axis is parallel to the rotation axis of Earth
and the $X-Y$ plane is same as the equatorial plane. 
Let us assume that the preferred axis $z$ in this frame can be expressed as,
\begin{equation}
\hat z = \cos\theta_p\hat Z+ \sin\theta_p\left(\cos\phi_p \hat X + \sin\phi_p
\hat Y \right)
\end{equation}
The axis $y''$ makes an angle $(\pi/2)-\lambda$ with respect to the $Z$ axis
at all times.
At some initial time $t=0$ let the azimuthal angle of $y''$ in this system be
$\alpha$.  
Hence we can express the laboratory frame $x''y''z''$ in terms of the
fixed coordinate system as
\begin{eqnarray}
\hat y'' &=& \sin\lambda\hat Z + \cos\lambda\left(\cos\alpha \hat X +
\sin\alpha \hat Y\right)\nonumber\\
\hat z'' &=& -\cos\lambda\hat Z + \sin\lambda\left(\cos\alpha \hat X +
\sin\alpha \hat Y\right)\nonumber\\
\hat x'' &=& -\sin\alpha \hat X + \cos\alpha \hat Y
\end{eqnarray}
At a later time $t$ the same formulas hold with the angle $\alpha$ 
replaced by $\tilde\alpha=\alpha+\delta$, where $\delta = 2\pi t/t_0$ and
$t_0$ is equal to a sidereal day. 
Using this we can directly compute the angles $\theta$ and $\beta$ at any time
by using $\cos\theta = \hat z\cdot \hat z''$, $\sin\theta\cos\beta = 
(\hat z\times \hat z'')\cdot \hat y''$ and
$\sin\theta\sin\beta = 
-(\hat z\times \hat z'')\cdot \hat x''$. Here $0\le\theta\le \pi$ and
$0\le \beta<2\pi$.  
The time dependences of $\theta$ and $\beta$ are shown in Fig. \ref{fig:tdep}
for a particular choice of parameters $\lambda$, $\theta_p$, $\phi_p$ and
$\alpha$. 

The daily variation of differential decay rate provides a very interesting
way to test the LV contribution due to VSR. We may divide
each sidereal day into a chosen number of bins. The data in each bin
can be accummulated over a large number of days in order to test for
the daily variation in the peak position of the azimuthal ($\phi''$) 
distribution. 
Correspondingly we can test the time dependence of the $\theta'$ 
(or $\theta''$) of the
decay rate. Here $\theta'$ (or $\theta''$) 
is simply the angle of the muon momentum
relative of the beam direction. In testing the angular dependence the main
complication is the detector response, which may not be isotropic. However
the detector response is not expected to be time dependent. Hence it
can be removed by subtracting out the time independent component in the 
$\phi''$ and $\theta'$ distributions.

\section{Conclusion}
We have studied several phenomenological implication of VSR starting
from an effective action approach in which we assume that the VSR term
acts as a small perturbation to the Standard Model action. The Lorentz 
violating VSR invariant terms are interesting since they may lead to
neutrino masses and mixing without requiring a right handed neutrino. 
Although this is possible we find that the resulting model becomes
intractable due to the non-diagonal nature of the resulting charged lepton
VSR mass matrix. The problem arises since the model, in general, 
does not admit a unitary evolution operator. We then impose some
constraints on the VSR mass parameters so that this problem does not
arise and we can reliably
determine its phenomenological implications. This requires us to
set VSR mass $M_-^2=0$ and furthermore assume that $M^2_+$ is diagonal
both for quarks and leptons.     

We determine the limits that can be imposed by the torsion pendulum
experiment and the clock comparison experiment on the VSR parameters. 
It is generally expected that Lorentz violation will lead to a
periodic time varying signal in these experiments with a period
of 1 sidereal day. Extensive searches for such signals have lead 
to null results \cite{Heckel:2006,Hughes:1960,Prestage:1985,Berglund:1995,Kostelecky:1999mr,Bear:2000,Phillips:2001,Cane:2004}  
We find that VSR also predicts a time depend signal in such experiments,
however the signal shows two complete oscillations with varying 
amplitude over a period of one sidereal day. 
Hence it is not possible to impose reliable limits on the VSR parameters
directly from the limits obtained by assuming a generic Lorentz violating
model.
A dedicated search is required which may pursued in future. 
We determine the level at which the VSR parameters for electron and 
nucleon (or up and down quarks) can be constrained by such experiments. 

Finally we study the implications of VSR in elementary particle experiments 
 by considering the charged pion and kaon decay processes,  
$\pi^-(q) \rightarrow \mu^-(p) + \bar{\nu}_{\mu}(k)$ and
$K^-(q)\rightarrow \mu^-(p) +\bar{\nu}_\mu(k)$ respectively. 
We impose a limit
on the VSR contributions to the up, down and strange quark masses by
using the known uncertainty in the decay rate of these processes. 
A more stringent limit may be imposed by studying the angular distribution 
of the decay products. 
 Due to the presence of a preferred direction in VSR, we
find that final state muon distribution acquires an azimuthal angle 
dependence relative to pion (or kaon) beam. Furthermore both the azimuthal and 
polar angle distributions acquire periodic time dependence with a 
period of one sidereal day. This time dependence provides us with an
effective way to test the principle of VSR at future particle physics 
experiments. The phenomenon is not limited to pion (or kaon) decay but may be
observed in many decay and scattering processes if VSR is the true symmetry
of nature. Excluding electron, up and down quarks, the most stringent limits
on the VSR contribution to fermion masses is expected to arise from
elementary particle physics experiments. 
 Furthermore
the phenomenon is different from the LV induced by quantum gravity effects
\cite{Colladay:1998fq,Collins:2004bp,GrootNibbelink:2004za,Jain:2005as,Polchinski:2011za} 
and might be observable at energies accessible in current or future colliders.

\bibliographystyle{apsrev}
\bibliography{Ref_muon}

\begin{thebibliography}{35}
\expandafter\ifx\csname natexlab\endcsname\relax\def\natexlab#1{#1}\fi
\expandafter\ifx\csname bibnamefont\endcsname\relax
  \def\bibnamefont#1{#1}\fi
\expandafter\ifx\csname bibfnamefont\endcsname\relax
  \def\bibfnamefont#1{#1}\fi
\expandafter\ifx\csname citenamefont\endcsname\relax
  \def\citenamefont#1{#1}\fi
\expandafter\ifx\csname url\endcsname\relax
  \def\url#1{\texttt{#1}}\fi
\expandafter\ifx\csname urlprefix\endcsname\relax\def\urlprefix{URL }\fi
\providecommand{\bibinfo}[2]{#2}
\providecommand{\eprint}[2][]{\url{#2}}

\bibitem[{\citenamefont{Colladay and Kostelecky}(1998)}]{Colladay:1998fq}
\bibinfo{author}{\bibfnamefont{D.}~\bibnamefont{Colladay}} \bibnamefont{and}
  \bibinfo{author}{\bibfnamefont{V.~A.} \bibnamefont{Kostelecky}},
  \bibinfo{journal}{Phys. Rev.} \textbf{\bibinfo{volume}{D58}},
  \bibinfo{pages}{116002} (\bibinfo{year}{1998}), \eprint{hep-ph/9809521}.

\bibitem[{\citenamefont{Collins et~al.}(2004)\citenamefont{Collins, Perez,
  Sudarsky, Urrutia, and Vucetich}}]{Collins:2004bp}
\bibinfo{author}{\bibfnamefont{J.}~\bibnamefont{Collins}},
  \bibinfo{author}{\bibfnamefont{A.}~\bibnamefont{Perez}},
  \bibinfo{author}{\bibfnamefont{D.}~\bibnamefont{Sudarsky}},
  \bibinfo{author}{\bibfnamefont{L.}~\bibnamefont{Urrutia}}, \bibnamefont{and}
  \bibinfo{author}{\bibfnamefont{H.}~\bibnamefont{Vucetich}},
  \bibinfo{journal}{Phys. Rev. Lett.} \textbf{\bibinfo{volume}{93}},
  \bibinfo{pages}{191301} (\bibinfo{year}{2004}), \eprint{gr-qc/0403053}.

\bibitem[{\citenamefont{Groot~Nibbelink and
  Pospelov}(2005)}]{GrootNibbelink:2004za}
\bibinfo{author}{\bibfnamefont{S.}~\bibnamefont{Groot~Nibbelink}}
  \bibnamefont{and} \bibinfo{author}{\bibfnamefont{M.}~\bibnamefont{Pospelov}},
  \bibinfo{journal}{Phys. Rev. Lett.} \textbf{\bibinfo{volume}{94}},
  \bibinfo{pages}{081601} (\bibinfo{year}{2005}), \eprint{hep-ph/0404271}.

\bibitem[{\citenamefont{Jain and Ralston}(2005)}]{Jain:2005as}
\bibinfo{author}{\bibfnamefont{P.}~\bibnamefont{Jain}} \bibnamefont{and}
  \bibinfo{author}{\bibfnamefont{J.~P.} \bibnamefont{Ralston}},
  \bibinfo{journal}{Phys. Lett.} \textbf{\bibinfo{volume}{B621}},
  \bibinfo{pages}{213} (\bibinfo{year}{2005}), \eprint{hep-ph/0502106}.

\bibitem[{\citenamefont{Polchinski}(2012)}]{Polchinski:2011za}
\bibinfo{author}{\bibfnamefont{J.}~\bibnamefont{Polchinski}},
  \bibinfo{journal}{Class. Quant. Grav.} \textbf{\bibinfo{volume}{29}},
  \bibinfo{pages}{088001} (\bibinfo{year}{2012}), \eprint{1106.6346}.

\bibitem[{\citenamefont{Cohen and Glashow}(2006{\natexlab{a}})}]{Cohen:2006ky}
\bibinfo{author}{\bibfnamefont{A.~G.} \bibnamefont{Cohen}} \bibnamefont{and}
  \bibinfo{author}{\bibfnamefont{S.~L.} \bibnamefont{Glashow}},
  \bibinfo{journal}{Phys. Rev. Lett.} \textbf{\bibinfo{volume}{97}},
  \bibinfo{pages}{021601} (\bibinfo{year}{2006}{\natexlab{a}}),
  \eprint{hep-ph/0601236}.

\bibitem[{\citenamefont{Das and Mohanty}(2011)}]{Das:2009fi}
\bibinfo{author}{\bibfnamefont{S.}~\bibnamefont{Das}} \bibnamefont{and}
  \bibinfo{author}{\bibfnamefont{S.}~\bibnamefont{Mohanty}},
  \bibinfo{journal}{Mod. Phys. Lett.} \textbf{\bibinfo{volume}{A26}},
  \bibinfo{pages}{139} (\bibinfo{year}{2011}), \eprint{0902.4549}.

\bibitem[{\citenamefont{Cohen and Glashow}(2006{\natexlab{b}})}]{Cohen:2006ir}
\bibinfo{author}{\bibfnamefont{A.~G.} \bibnamefont{Cohen}} \bibnamefont{and}
  \bibinfo{author}{\bibfnamefont{S.~L.} \bibnamefont{Glashow}}
  (\bibinfo{year}{2006}{\natexlab{b}}), \eprint{hep-ph/0605036}.

\bibitem[{\citenamefont{{Dunn} and {Mehen}}(2006)}]{Dunn:2006}
\bibinfo{author}{\bibfnamefont{A.}~\bibnamefont{{Dunn}}} \bibnamefont{and}
  \bibinfo{author}{\bibfnamefont{T.}~\bibnamefont{{Mehen}}},
  \bibinfo{journal}{ArXiv High Energy Physics - Phenomenology e-prints}
  (\bibinfo{year}{2006}), \eprint{hep-ph/0610202}.

\bibitem[{\citenamefont{Fan et~al.}(2007)\citenamefont{Fan, Goldberger, and
  Skiba}}]{Fan:2006nd}
\bibinfo{author}{\bibfnamefont{J.}~\bibnamefont{Fan}},
  \bibinfo{author}{\bibfnamefont{W.~D.} \bibnamefont{Goldberger}},
  \bibnamefont{and} \bibinfo{author}{\bibfnamefont{W.}~\bibnamefont{Skiba}},
  \bibinfo{journal}{Phys. Lett.} \textbf{\bibinfo{volume}{B649}},
  \bibinfo{pages}{186} (\bibinfo{year}{2007}), \eprint{hep-ph/0611049}.

\bibitem[{\citenamefont{Cohen and Freedman}(2007)}]{Cohen:2006sc}
\bibinfo{author}{\bibfnamefont{A.~G.} \bibnamefont{Cohen}} \bibnamefont{and}
  \bibinfo{author}{\bibfnamefont{D.~Z.} \bibnamefont{Freedman}},
  \bibinfo{journal}{JHEP} \textbf{\bibinfo{volume}{07}}, \bibinfo{pages}{039}
  (\bibinfo{year}{2007}), \eprint{hep-th/0605172}.

\bibitem[{\citenamefont{Gibbons et~al.}(2007)\citenamefont{Gibbons, Gomis, and
  Pope}}]{Gibbons:2007iu}
\bibinfo{author}{\bibfnamefont{G.~W.} \bibnamefont{Gibbons}},
  \bibinfo{author}{\bibfnamefont{J.}~\bibnamefont{Gomis}}, \bibnamefont{and}
  \bibinfo{author}{\bibfnamefont{C.~N.} \bibnamefont{Pope}},
  \bibinfo{journal}{Phys. Rev.} \textbf{\bibinfo{volume}{D76}},
  \bibinfo{pages}{081701} (\bibinfo{year}{2007}), \eprint{0707.2174}.

\bibitem[{\citenamefont{Bernardini and da~Rocha}(2008)}]{Bernardini:2007ex}
\bibinfo{author}{\bibfnamefont{A.~E.} \bibnamefont{Bernardini}}
  \bibnamefont{and} \bibinfo{author}{\bibfnamefont{R.}~\bibnamefont{da~Rocha}},
  \bibinfo{journal}{Europhys. Lett.} \textbf{\bibinfo{volume}{81}},
  \bibinfo{pages}{40010} (\bibinfo{year}{2008}), \eprint{hep-th/0701092}.

\bibitem[{\citenamefont{Lee}(2016)}]{Lee:2015tcc}
\bibinfo{author}{\bibfnamefont{C.-Y.} \bibnamefont{Lee}},
  \bibinfo{journal}{Phys. Rev.} \textbf{\bibinfo{volume}{D93}},
  \bibinfo{pages}{045011} (\bibinfo{year}{2016}), \eprint{1512.09175}.

\bibitem[{\citenamefont{Sheikh-Jabbari and
  Tureanu}(2008)}]{SheikhJabbari:2008nc}
\bibinfo{author}{\bibfnamefont{M.~M.} \bibnamefont{Sheikh-Jabbari}}
  \bibnamefont{and} \bibinfo{author}{\bibfnamefont{A.}~\bibnamefont{Tureanu}},
  \bibinfo{journal}{Phys. Rev. Lett.} \textbf{\bibinfo{volume}{101}},
  \bibinfo{pages}{261601} (\bibinfo{year}{2008}), \eprint{0806.3699}.

\bibitem[{\citenamefont{Ahluwalia and Horvath}(2010)}]{Ahluwalia:2010zn}
\bibinfo{author}{\bibfnamefont{D.~V.} \bibnamefont{Ahluwalia}}
  \bibnamefont{and} \bibinfo{author}{\bibfnamefont{S.~P.}
  \bibnamefont{Horvath}}, \bibinfo{journal}{JHEP}
  \textbf{\bibinfo{volume}{11}}, \bibinfo{pages}{078} (\bibinfo{year}{2010}),
  \eprint{1008.0436}.

\bibitem[{\citenamefont{{Voh{\'a}nka}}(2012)}]{2012PhRvD..85j5009V}
\bibinfo{author}{\bibfnamefont{J.}~\bibnamefont{{Voh{\'a}nka}}},
  \bibinfo{journal}{\prd} \textbf{\bibinfo{volume}{85}}, \bibinfo{eid}{105009}
  (\bibinfo{year}{2012}), \eprint{1112.1797}.

\bibitem[{\citenamefont{Alfaro and Rivelles}(2013)}]{Alfaro:2013uva}
\bibinfo{author}{\bibfnamefont{J.}~\bibnamefont{Alfaro}} \bibnamefont{and}
  \bibinfo{author}{\bibfnamefont{V.~O.} \bibnamefont{Rivelles}},
  \bibinfo{journal}{Phys.Rev.} \textbf{\bibinfo{volume}{D88}},
  \bibinfo{pages}{085023} (\bibinfo{year}{2013}), \eprint{1305.1577}.

\bibitem[{\citenamefont{Cheon et~al.}(2009)\citenamefont{Cheon, Lee, and
  Lee}}]{Cheon:2009zx}
\bibinfo{author}{\bibfnamefont{S.}~\bibnamefont{Cheon}},
  \bibinfo{author}{\bibfnamefont{C.}~\bibnamefont{Lee}}, \bibnamefont{and}
  \bibinfo{author}{\bibfnamefont{S.~J.} \bibnamefont{Lee}},
  \bibinfo{journal}{Phys. Lett.} \textbf{\bibinfo{volume}{B679}},
  \bibinfo{pages}{73} (\bibinfo{year}{2009}), \eprint{0904.2065}.

\bibitem[{\citenamefont{{Alfaro} et~al.}(2015)\citenamefont{{Alfaro},
  {Gonz{\'a}lez}, and {{\'A}vila}}}]{Alfaro:2015fha}
\bibinfo{author}{\bibfnamefont{J.}~\bibnamefont{{Alfaro}}},
  \bibinfo{author}{\bibfnamefont{P.}~\bibnamefont{{Gonz{\'a}lez}}},
  \bibnamefont{and}
  \bibinfo{author}{\bibfnamefont{R.}~\bibnamefont{{{\'A}vila}}},
  \bibinfo{journal}{Phys. Rev.} \textbf{\bibinfo{volume}{D91}},
  \bibinfo{pages}{105007} (\bibinfo{year}{2015}), \bibinfo{note}{[Addendum:
  Phys. Rev.D91,no.12,129904(2015)]}, \eprint{1504.04222}.

\bibitem[{\citenamefont{Heckel et~al.}(2006)\citenamefont{Heckel, Cramer, Cook,
  Adelberger, Schlamminger, and Schmidt}}]{Heckel:2006}
\bibinfo{author}{\bibfnamefont{B.~R.} \bibnamefont{Heckel}},
  \bibinfo{author}{\bibfnamefont{C.~E.} \bibnamefont{Cramer}},
  \bibinfo{author}{\bibfnamefont{T.~S.} \bibnamefont{Cook}},
  \bibinfo{author}{\bibfnamefont{E.~G.} \bibnamefont{Adelberger}},
  \bibinfo{author}{\bibfnamefont{S.}~\bibnamefont{Schlamminger}},
  \bibnamefont{and} \bibinfo{author}{\bibfnamefont{U.}~\bibnamefont{Schmidt}},
  \bibinfo{journal}{Phys. Rev. Lett.} \textbf{\bibinfo{volume}{97}},
  \bibinfo{pages}{021603} (\bibinfo{year}{2006}).

\bibitem[{\citenamefont{Can\`e et~al.}(2004)\citenamefont{Can\`e, Bear,
  Phillips, Rosen, Smallwood, Stoner, Walsworth, and Kosteleck\'y}}]{Cane:2004}
\bibinfo{author}{\bibfnamefont{F.}~\bibnamefont{Can\`e}},
  \bibinfo{author}{\bibfnamefont{D.}~\bibnamefont{Bear}},
  \bibinfo{author}{\bibfnamefont{D.~F.} \bibnamefont{Phillips}},
  \bibinfo{author}{\bibfnamefont{M.~S.} \bibnamefont{Rosen}},
  \bibinfo{author}{\bibfnamefont{C.~L.} \bibnamefont{Smallwood}},
  \bibinfo{author}{\bibfnamefont{R.~E.} \bibnamefont{Stoner}},
  \bibinfo{author}{\bibfnamefont{R.~L.} \bibnamefont{Walsworth}},
  \bibnamefont{and} \bibinfo{author}{\bibfnamefont{V.~A.}
  \bibnamefont{Kosteleck\'y}}, \bibinfo{journal}{Phys. Rev. Lett.}
  \textbf{\bibinfo{volume}{93}}, \bibinfo{pages}{230801}
  (\bibinfo{year}{2004}).

\bibitem[{\citenamefont{Garg et~al.}(2011)\citenamefont{Garg, Shreecharan, Das,
  Deshpande, and Rajasekaran}}]{Garg:2011aa}
\bibinfo{author}{\bibfnamefont{S.~K.} \bibnamefont{Garg}},
  \bibinfo{author}{\bibfnamefont{T.}~\bibnamefont{Shreecharan}},
  \bibinfo{author}{\bibfnamefont{P.~K.} \bibnamefont{Das}},
  \bibinfo{author}{\bibfnamefont{N.~G.} \bibnamefont{Deshpande}},
  \bibnamefont{and}
  \bibinfo{author}{\bibfnamefont{G.}~\bibnamefont{Rajasekaran}},
  \bibinfo{journal}{JHEP} \textbf{\bibinfo{volume}{07}}, \bibinfo{pages}{024}
  (\bibinfo{year}{2011}), \eprint{1105.5203}.

\bibitem[{\citenamefont{Altschul}(2013)}]{Altschul:2013yja}
\bibinfo{author}{\bibfnamefont{B.}~\bibnamefont{Altschul}},
  \bibinfo{journal}{Phys. Rev.} \textbf{\bibinfo{volume}{D88}},
  \bibinfo{pages}{076015} (\bibinfo{year}{2013}), \eprint{1308.2602}.

\bibitem[{\citenamefont{Noordmans and Vos}(2014)}]{Noordmans:2014bua}
\bibinfo{author}{\bibfnamefont{J.~P.} \bibnamefont{Noordmans}}
  \bibnamefont{and} \bibinfo{author}{\bibfnamefont{K.~K.} \bibnamefont{Vos}},
  \bibinfo{journal}{Phys. Rev.} \textbf{\bibinfo{volume}{D89}},
  \bibinfo{pages}{101702} (\bibinfo{year}{2014}), \eprint{1404.7629}.

\bibitem[{\citenamefont{Mittleman et~al.}(1999)\citenamefont{Mittleman,
  Ioannou, Dehmelt, and Russell}}]{Mittleman:1999}
\bibinfo{author}{\bibfnamefont{R.~K.} \bibnamefont{Mittleman}},
  \bibinfo{author}{\bibfnamefont{I.~I.} \bibnamefont{Ioannou}},
  \bibinfo{author}{\bibfnamefont{H.~G.} \bibnamefont{Dehmelt}},
  \bibnamefont{and} \bibinfo{author}{\bibfnamefont{N.}~\bibnamefont{Russell}},
  \bibinfo{journal}{Phys. Rev. Lett.} \textbf{\bibinfo{volume}{83}},
  \bibinfo{pages}{2116} (\bibinfo{year}{1999}).

\bibitem[{\citenamefont{Hughes et~al.}(1960)\citenamefont{Hughes, Robinson, and
  Beltran-Lopez}}]{Hughes:1960}
\bibinfo{author}{\bibfnamefont{V.~W.} \bibnamefont{Hughes}},
  \bibinfo{author}{\bibfnamefont{H.~G.} \bibnamefont{Robinson}},
  \bibnamefont{and}
  \bibinfo{author}{\bibfnamefont{V.}~\bibnamefont{Beltran-Lopez}},
  \bibinfo{journal}{Phys. Rev. Lett.} \textbf{\bibinfo{volume}{4}},
  \bibinfo{pages}{342} (\bibinfo{year}{1960}).

\bibitem[{\citenamefont{Prestage et~al.}(1985)\citenamefont{Prestage,
  Bollinger, Itano, and Wineland}}]{Prestage:1985}
\bibinfo{author}{\bibfnamefont{J.~D.} \bibnamefont{Prestage}},
  \bibinfo{author}{\bibfnamefont{J.~J.} \bibnamefont{Bollinger}},
  \bibinfo{author}{\bibfnamefont{W.~M.} \bibnamefont{Itano}}, \bibnamefont{and}
  \bibinfo{author}{\bibfnamefont{D.~J.} \bibnamefont{Wineland}},
  \bibinfo{journal}{Phys. Rev. Lett.} \textbf{\bibinfo{volume}{54}},
  \bibinfo{pages}{2387} (\bibinfo{year}{1985}).

\bibitem[{\citenamefont{Berglund et~al.}(1995)\citenamefont{Berglund, Hunter,
  Krause, Prigge, Ronfeldt, and Lamoreaux}}]{Berglund:1995}
\bibinfo{author}{\bibfnamefont{C.~J.} \bibnamefont{Berglund}},
  \bibinfo{author}{\bibfnamefont{L.~R.} \bibnamefont{Hunter}},
  \bibinfo{author}{\bibfnamefont{D.}~\bibnamefont{Krause}, \bibfnamefont{Jr.}},
  \bibinfo{author}{\bibfnamefont{E.~O.} \bibnamefont{Prigge}},
  \bibinfo{author}{\bibfnamefont{M.~S.} \bibnamefont{Ronfeldt}},
  \bibnamefont{and} \bibinfo{author}{\bibfnamefont{S.~K.}
  \bibnamefont{Lamoreaux}}, \bibinfo{journal}{Phys. Rev. Lett.}
  \textbf{\bibinfo{volume}{75}}, \bibinfo{pages}{1879} (\bibinfo{year}{1995}).

\bibitem[{\citenamefont{Kostelecky and Lane}(1999)}]{Kostelecky:1999mr}
\bibinfo{author}{\bibfnamefont{V.~A.} \bibnamefont{Kostelecky}}
  \bibnamefont{and} \bibinfo{author}{\bibfnamefont{C.~D.} \bibnamefont{Lane}},
  \bibinfo{journal}{Phys. Rev.} \textbf{\bibinfo{volume}{D60}},
  \bibinfo{pages}{116010} (\bibinfo{year}{1999}), \eprint{hep-ph/9908504}.

\bibitem[{\citenamefont{Bear et~al.}(2000)\citenamefont{Bear, Stoner,
  Walsworth, Kosteleck\'y, and Lane}}]{Bear:2000}
\bibinfo{author}{\bibfnamefont{D.}~\bibnamefont{Bear}},
  \bibinfo{author}{\bibfnamefont{R.~E.} \bibnamefont{Stoner}},
  \bibinfo{author}{\bibfnamefont{R.~L.} \bibnamefont{Walsworth}},
  \bibinfo{author}{\bibfnamefont{V.~A.} \bibnamefont{Kosteleck\'y}},
  \bibnamefont{and} \bibinfo{author}{\bibfnamefont{C.~D.} \bibnamefont{Lane}},
  \bibinfo{journal}{Phys. Rev. Lett.} \textbf{\bibinfo{volume}{85}},
  \bibinfo{pages}{5038} (\bibinfo{year}{2000}).

\bibitem[{\citenamefont{Phillips et~al.}(2001)\citenamefont{Phillips, Humphrey,
  Mattison, Stoner, Vessot, and Walsworth}}]{Phillips:2001}
\bibinfo{author}{\bibfnamefont{D.~F.} \bibnamefont{Phillips}},
  \bibinfo{author}{\bibfnamefont{M.~A.} \bibnamefont{Humphrey}},
  \bibinfo{author}{\bibfnamefont{E.~M.} \bibnamefont{Mattison}},
  \bibinfo{author}{\bibfnamefont{R.~E.} \bibnamefont{Stoner}},
  \bibinfo{author}{\bibfnamefont{R.~F.~C.} \bibnamefont{Vessot}},
  \bibnamefont{and} \bibinfo{author}{\bibfnamefont{R.~L.}
  \bibnamefont{Walsworth}}, \bibinfo{journal}{Phys. Rev. D}
  \textbf{\bibinfo{volume}{63}}, \bibinfo{pages}{111101}
  (\bibinfo{year}{2001}).

\bibitem[{\citenamefont{Jain and Munczek}(1993)}]{Jain:1993qh}
\bibinfo{author}{\bibfnamefont{P.}~\bibnamefont{Jain}} \bibnamefont{and}
  \bibinfo{author}{\bibfnamefont{H.~J.} \bibnamefont{Munczek}},
  \bibinfo{journal}{Phys. Rev.} \textbf{\bibinfo{volume}{D48}},
  \bibinfo{pages}{5403} (\bibinfo{year}{1993}), \eprint{hep-ph/9307221}.

\bibitem[{\citenamefont{Patrignani et~al.}(2016)}]{Patrignani:2016xqp}
\bibinfo{author}{\bibfnamefont{C.}~\bibnamefont{Patrignani}}
  \bibnamefont{et~al.} (\bibinfo{collaboration}{Particle Data Group}),
  \bibinfo{journal}{Chin. Phys.} \textbf{\bibinfo{volume}{C40}},
  \bibinfo{pages}{100001} (\bibinfo{year}{2016}).

\bibitem[{\citenamefont{Capstick and Roberts}(2000)}]{Capstick:2000qj}
\bibinfo{author}{\bibfnamefont{S.}~\bibnamefont{Capstick}} \bibnamefont{and}
  \bibinfo{author}{\bibfnamefont{W.}~\bibnamefont{Roberts}},
  \bibinfo{journal}{Prog. Part. Nucl. Phys.} \textbf{\bibinfo{volume}{45}},
  \bibinfo{pages}{S241} (\bibinfo{year}{2000}), \eprint{nucl-th/0008028}.

\end{thebibliography}

\end{document}